# Dynamical Symmetry Breaking by SU(2) Gauge Bosons

F.J. Himpsel

Department of Physics, University of Wisconsin Madison,
1150 University Ave., Madison, WI 53706, USA, fhimpsel@wisc.edu

**Abstract**

This work explores the possibility of obtaining a mass gap in Yang-Mills theories via the intrinsic gauge bosons, without invoking a separate Higgs boson or fermion‐antifermion pairs. Instead, pairs of gauge bosons in the spin and isospin singlet state form a pair of composite Higgs bosons which can be viewed as the simplest possible glueball of Yang-Mills gauge theories. Quadratic and quartic gauge boson self-interactions form a potential that leads to a finite expectation value of the gauge boson amplitude. Transverse polarization ensures Lorentz invariance of the vacuum after averaging over all possible polarization vectors. But the scalar pair products exhibit a finite vacuum expectation value which breaks the gauge symmetry dynamically. Compatibility with the standard Higgs potential determines the quadratic and quartic coupling constants.

# 1. Introduction

Calculating adjustable parameters of the standard model from first principles has been a long-time challenge. Particularly mysterious have been the two parameters determining the Higgs potential of the standard model. One of them (labeled $-\mu^2$) corresponds to an imaginary mass and the other (labeled $\lambda$) belongs to a quartic Lagrangian. None of the other fundamental particles exhibits such a Lagrangian. This *ad-hoc* potential is responsible for breaking the SU(2)×U(1)$_Y$ gauge symmetry of the electroweak interaction by creating a finite vacuum expectation value (VEV) for the Higgs boson. That in turn conveys mass to fundamental particles.

Such considerations led to models where the Higgs boson is not fundamental, but composite [1],[2]. In most cases the constituents were fermion-antifermion pairs [1], but a Higgs boson composed of the three SU(2) gauge bosons was proposed as well [2]. It explained the Higgs mass, which became simply half of the standard Higgs VEV in lowest order. That matched the experimental result [3] with tree-level accuracy ($\approx 2\%$).

Apart from exploring the origin of mass in particle physics, the SU(2) gauge theory has attracted interest in mathematical physics [4]. The three SU(2) gauge bosons form the simplest non-abelian gauge theory. Such Yang-Mills theories play a dominant role in the standard model and its extensions. A particular concern has been the very existence of such theories (by rigorous mathematical standards), together with the mechanism of dynamical symmetry breaking and the resulting mass gap [4].

In the following we start out with a review of the composite Higgs model proposed in [2], restricted to the minimal set of particles, i.e., the three gauge bosons of the SU(2) Yang-Mills theory. The definition of the composite Higgs boson from the gauge bosons is worked out and its consequence on the Higgs mass is demonstrated. This section establishes several relations between the composite Higgs boson and the SU(2) gauge bosons. Section 3 investigates the precise form of the expectation value (EV) for the amplitude of a gauge field, as dictated by Lorentz and gauge invariance. The gauge fields in the Lagrangian are decomposed into their EVs and observable gauge bosons. Section 4 introduces the gauge boson potential which breaks the SU(2) gauge symmetry and creates VEVs for gauge boson pairs. Gauge-invariant model Lagrangians serve as building blocks for this potential. Section 5 exploits compatibility criteria between the potentials of the gauge bosons and the Higgs boson. Those fix the quadratic and quartic coupling constants. Section 6 summarizes the concept of dynamical symmetry breaking by gauge bosons and places it into the broader context of Yang-Mills theories. Appendix A discusses tree-level gauge boson self-interactions and explains why they do not contribute to the symmetry-breaking potential. Appendix B outlines the extension of the SU(2) model to the electroweak SU(2)×U(1)$_Y$ symmetry.

**2. Composite Higgs Model for Pure SU(2) Gauge Symmetry**

The pure SU(2) model is chosen to provide clearer insight into the concept of a Higgs boson composed of gauge bosons which was developed originally for the full SU(2)×U(1)$_Y$ electroweak symmetry in [2]. Mixing with the U(1)$_Y$ hypercharge symmetry does not affect the $W^1, W^2$ gauge bosons that form the observed $W^{\pm}$ particle, leaving their mass and couplings unchanged. The consequences of electroweak mixing will be addressed briefly in Appendix B.

The strategy for replacing the Higgs boson of the standard model by a composite of SU(2) gauge bosons can be summarized as follows:

1) Remove the Higgs boson from the Lagrangian.
2) Replace it by a Lorentz- and gauge-invariant composite of SU(2) gauge bosons.
3) Establish a potential for the gauge bosons from their self-interactions.
4) Obtain EVs for the gauge bosons and VEVs for their scalar products by generalizing the Brout-Englert-Higgs mechanism from scalars to vectors.
5) Transfer VEVs and masses from the gauge bosons to the composite Higgs boson.

The standard Higgs field can be written as the combination of a SU(2) singlet $H_0$ with a triplet of Nambu-Goldstone modes $(w_1, w_2, w_3)$, forming a complex doublet $\Phi_0$. The subscript 0 is used to label fields appearing in the original, gauge-invariant Lagrangian. These are decomposed into a VEV and an observable ("physical") field. The Higgs field $\Phi_0$ can be represented by a 2×2 matrix $\mathbf{\Phi}_0$ using the 2×2 unit matrix **1** and the Pauli matrices $\tau^j$ (with 2×2 matrices in bold):

(1) $\quad \mathbf{\Phi}_0 = \tfrac{1}{\sqrt{2}}(H_0 \cdot \mathbf{1} + i\, \Sigma_j w_j \cdot \boldsymbol{\tau}^j) \qquad \Phi_0 = \mathbf{\Phi}_0 \cdot \begin{bmatrix} 0 \\ 1 \end{bmatrix} = \tfrac{1}{\sqrt{2}} \begin{bmatrix} w_2 + i w_1 \\ H_0 - i w_3 \end{bmatrix} \qquad \Phi_0^C = \mathbf{\Phi}_0 \cdot \begin{bmatrix} 1 \\ 0 \end{bmatrix}$

$$H_0 = \langle H_0 \rangle + H \qquad \langle H_0 \rangle = v = 2^{-1/4} G_F^{-1/2} = 246.22 \text{ GeV} \qquad \langle w_j \rangle = 0$$

$\Phi_0^C$ is the charge conjugate of $\Phi_0$. The singlet acquires a finite VEV $\langle H_0 \rangle = v$ via the Brout-Englert-Higgs mechanism. Its value is directly related to the experimental four-fermion coupling constant $G_F$ [3]. The VEVs of the Goldstone modes vanish.

The standard Higgs potential combines a quadratic with a biquadratic term:

(2) $\quad V_\Phi = -\mu^2 \cdot \Phi_0^\dagger \Phi_0 + \lambda \cdot [\Phi_0^\dagger \Phi_0]^2 \qquad\qquad$ General gauge

$\qquad V_\Phi = -½\mu^2 \cdot H_0^2 + ¼\lambda \cdot H_0^4 \qquad\qquad$ Unitary gauge

Using the pairs $\Phi_0^\dagger \Phi_0$ or $H_0^2$ as variables simplifies the potential to a linear plus a quadratic term, providing a hint that pairs may play a role in Higgs interactions.

The SU(2) gauge bosons form a triplet $(W_0^1, W_0^2, W_0^3)$. The sum over gauge boson pairs $\Sigma_i W_{0,\mu}^i W_0^{i,\mu}$ is a Lorentz scalar and a SU(2) singlet, thereby matching $\Phi_0^\dagger \Phi_0$. That suggests a proportionality between a pair of Higgs bosons and pairs of gauge bosons:

(3) $\quad \Phi_0^\dagger \Phi_0 = ½[H_0^2 + \Sigma_i w_i^2] \propto -½ \Sigma_i W_{0,\mu}^i W_0^{i,\mu} = -½ \Sigma_i (W_0^i W_0^i) \qquad$ (+---) metric

Scalar products of gauge bosons are abbreviated by round brackets. They are negative for the space-like gauge bosons.

Instead of defining $\Phi_0$ via (1), the Goldstones $w_i$ can be incorporated in nonlinear fashion as SU(2) matrix **U** (see [5] and references therein):

(4) $\quad \mathbf{U} = \exp\left[i \cdot \Sigma_j \frac{w_j}{v} \cdot \tau^j\right] = \cos\left[\frac{|w|}{v}\right] \cdot \mathbf{1} + i \cdot \left[\frac{|w|}{v}\right]^{-1} \sin\left[\frac{|w|}{v}\right] \cdot \Sigma_j \frac{w_j}{v} \cdot \tau^j \qquad |w| = (\Sigma_i w_i^2)^{½}$

$\qquad\qquad = \mathbf{1} + i \cdot \Sigma_j \frac{w_j}{v} \cdot \tau^j - ½\left[\frac{|w|}{v}\right]^2 \mathbf{1} - i \cdot \frac{1}{6}\left[\frac{|w|}{v}\right]^2 \cdot \Sigma_j \frac{w_j}{v} \cdot \tau^j + ...$

$\qquad \frac{1}{\sqrt{2}} H_0 \cdot \mathbf{U} \approx \mathbf{\Phi}_0 \qquad$ for $H, |w| \ll v$

The gauge bosons are incorporated via the gauge-invariant derivative $D_\mu$ of the matrix **U**:

(5) $\quad D_\mu \mathbf{U} = \partial_\mu \mathbf{U} - ig \cdot \mathbf{W}_{0,\mu} \mathbf{U} \qquad\qquad \mathbf{W}_{0,\mu} = \Sigma_j W_{0,\mu}^j \cdot ½\tau^j$

$D_\mu \mathbf{U}$ generates a four-vector of 2×2 matrices representing the gauge bosons $W_{0,\mu}^j$:

(6) $\quad \mathbf{V}_{0,\mu} = -\mathbf{V}_{0,\mu}^\dagger = [D_\mu \mathbf{U}] \mathbf{U}^\dagger = i \cdot \Sigma_j [\partial_\mu \frac{w_j}{v} - ½ g W_{0,\mu}^j] \cdot \tau^j + ...$

(7) $\quad \text{tr}[(\mathbf{V}_0 \mathbf{V}_0)] = -2 \Sigma_j [\partial_\mu \frac{w_j}{v} - ½ g W_{0,\mu}^j] \cdot [\partial^\mu \frac{w_j}{v} - ½ g W_0^{j,\mu}] + ... \qquad$ General gauge

The trace of $(\mathbf{V}_0 \mathbf{V}_0)$ provides a gauge-invariant definition of the composite Higgs boson:

(8) $\quad \boxed{\Phi_0^\dagger \Phi_0 = \text{tr}[(\mathbf{V}_0 \mathbf{V}_0)]} \qquad\qquad$ General gauge

(9) $\quad ½ H_0^2 + ½ \Sigma_i w_i^2 = -g^2 \cdot ½ \Sigma_i (W_0^i W_0^i) + g \cdot 2\Sigma_i \partial_\mu \frac{w_i}{v} \cdot W_0^{i,\mu} - 2\Sigma_i \partial_\mu \frac{w_i}{v} \cdot \partial^\mu \frac{w_i}{v} + ...$

(10) $\quad \boxed{½H_0^2 = -g^2 \cdot ½\Sigma_i (W_0^i W_0^i)}$ \hfill Unitary gauge

Notice a subtle, but conceptually-significant difference from Ref. [2], where the composite Higgs boson was defined in terms of observable gauge bosons. Here we use Lagrangian gauge bosons and derive the definition given in [2] from those in Section 3. Making the definition at the Lagrangian level preserves explicit SU(2) gauge symmetry.

The leading term on the left side of (9),(10) comes from the VEV of the Higgs boson $\langle H_0 \rangle^2 = v^2$, because the observable Higgs field H represents small oscillations about the VEV. Otherwise the VEV would not be noticed on top of the oscillations. This implies that the leading term on the right side must be due to finite VEVs of gauge boson pairs. A vector field with a finite VEV $\langle W_0^i \rangle$ would violate Lorentz invariance of the vacuum by specifying a specific direction in space-time. This problem is avoided by having different orientations for the expectation values $\langle W_0^i \rangle$ of individual field quanta, depending on their momenta [2]. These EVs change from one gauge boson to another and average out to zero after summing over all virtual gauge bosons in the vacuum of quantum field theory (compare the summation over the vacuum photons causing the Casimir effect). But the scalar products $(\langle W_0^i \rangle \langle W_0^i \rangle)$ are not averaged out. They provide a finite VEV to match that of the Higgs boson. The leading terms of (9),(10) then establish a relation between $v^2$ and scalar products of gauge boson EVs:

(11) $\quad \boxed{v^2 = -g^2 \cdot \Sigma_i (\langle W_0^i \rangle \langle W_0^i \rangle)}$ \hfill General gauge

$\qquad v^2 = 3g^2 w^2 \quad$ assuming $\ (\langle W_0^i \rangle \langle W_0^i \rangle) = -w^2 \quad$ for $\ i=1,2,3$

The higher order terms in (9),(10) will be worked out in Section 3 after pinpointing the structure of the gauge boson EVs. They lead to a relation between the observable Higgs boson H and the observable gauge bosons $W^i$:

(12) $\quad \boxed{\Phi^\dagger \Phi \approx \mathrm{tr}[(\mathbf{VV})]}$ \hfill General gauge

(13) $\quad ½H^2 + ½\Sigma_i w_i^2 \approx -g^2 \cdot ½\Sigma_i (W_T^i W_T^i) - 2\Sigma_i \partial^\mu \frac{w_i}{v} \cdot \partial_\mu \frac{w_i}{v} \qquad \partial^\mu \frac{w_i}{v} \to ½g W_L^{i,\mu}$

(14) $\quad \boxed{½H^2 \approx -g^2 \cdot ½\Sigma_i (W^i W^i)}$ \hfill Unitary gauge

For specific gauges one needs to separate the observable gauge bosons $W^i$ into transverse and longitudinal components ($W^i = W_T^i + W_L^i$). The last term in (13) can be assigned to longitudinal gauge bosons $W_L^{i,\mu}$. In the Landau gauge this term needs to be moved to the left side of (13), in order to obtain purely transverse gauge bosons.

An important consequence of (14) is the determination of the Higgs mass from the four-fermion coupling constant $G_F$ in [2]. Multiplication by $-(\frac{1}{2}v)^2$ reveals a connection between the mass Lagrangians of the observable Higgs and gauge bosons:

(15) $\quad L_M^H = -\frac{1}{2} M_H^2 \cdot H^2 \quad$ with $\quad \boxed{M_H \approx \frac{1}{2} v}$

(16) $\quad L_M^W = \frac{1}{2} M_W^2 \cdot \Sigma_i (W^i W^i) \quad$ with $\quad \boxed{M_W \approx \frac{1}{2} v g}$

$\Rightarrow L_M^H \approx L_M^W$

With the tree-level gauge boson mass $M_W \approx \frac{1}{2} v g$ from the standard model, the tree-level Higgs mass becomes $M_H \approx \frac{1}{2} v = 2^{-5/4} G_F^{-1/2} = 123.1$ GeV. That matches the experimental result of 125.1 GeV within 2% [3]. A similar match exists between $M_W \approx \frac{1}{2} v g = 77.5$ GeV and the observed value of 80.4 GeV. Such a margin is typical for the tree-level approximation which neglects corrections of the order $\alpha_w = g^2/4\pi \approx 3\%$.

### 3. Expectation Values of Gauge Bosons

The derivation of (12)-(14) from (8)-(10) calls for a more detailed analysis of the gauge boson EVs. They have to be transverse for two reasons: 1) Only the transverse modes are gauge-invariant, while the longitudinal mode is traded for a Goldstone scalar when going from the unitary gauge to the Landau gauge. 2) A gauge-symmetric Lagrangian requires massless gauge bosons which are purely transverse. These arguments also apply to the EVs. Therefore it is useful to choose the transverse Landau gauge for decomposing the Lagrangian gauge bosons $W_0^i$ into their EVs $\langle W_0^i \rangle$ and observable gauge bosons $W_T^i$. A subsequent transformation to the unitary gauge will remove the Goldstones, converting them to longitudinal gauge bosons $W_L^i$ (compare the discussion with (14)). Neither the Goldstones nor the longitudinal gauge bosons contribute an EV.

The definition of the composite Higgs boson involves pairs of gauge bosons. These orbit around their center of mass with opposite momenta, which serve as reference for their polarization vectors. The wave function of the singlet ground state has even parity with the spin configuration $(\uparrow\downarrow + \downarrow\uparrow)/\sqrt{2}$. The two gauge bosons have the same (circular) polarization in this state [6]. Thus one can assume that the three gauge bosons pairs defining the Higgs boson all have the same polarization $\varepsilon_\alpha$. In the (transverse) Landau gauge the observable gauge bosons become products of $\varepsilon_\alpha$ with scalar operators $w^i$ (distinct from the Goldstones $w_i$ and the VEVs $w^i$):

(17) $W_0^i = \langle W_0^i \rangle + W_T^i$    $\boxed{\langle W_0^i \rangle = w^i \cdot \varepsilon_\alpha}$    $\langle W_T^i \rangle = 0$    $\boxed{W_T^i = w^i \cdot \varepsilon_\alpha}$      Landau gauge

(18) $(\varepsilon_\alpha^* \varepsilon_\beta) = -\delta_{\alpha\beta}$      $\boxed{(W_T^i W_T^i) = -(w^i)^2}$

To obtain the relations (12)-(14) between the observable Higgs and gauge bosons we insert (17) into (9). The terms $\partial_\mu w_i \cdot W_0^{i,\mu}$ on the right side of (9) vanish, since the $W_0^i$ are transverse while the derivatives $i\partial_\mu w_i$ are parallel to the common four-momentum $k_\mu$:

(19) $\tfrac{1}{2} H_0^2 + \tfrac{1}{2} \Sigma_i w_i^2 = -g^2 \cdot \tfrac{1}{2} \Sigma_i (W_0^i W_0^i) - 2\Sigma_i \partial_\mu \tfrac{w_i}{v} \cdot \partial^\mu \tfrac{w_i}{v} + \ldots$      General gauge

    $\downarrow$                     $\downarrow$

$H_0^2 = v^2 + 2vH + H^2$      $(W_0^i W_0^i) = (\langle W_0^i \rangle \langle W_0^i \rangle) + 2 \cdot (\langle W_0^i \rangle W_T^i) + (W_T^i W_T^i)$

The conversions in the second line are based on (17), which is valid in the Landau gauge. The last term in (19) can be identified with the longitudinal gauge boson term $(W_L^i W_L^i)$, as discussed after (14). It supplements the transverse term $(W_T^i W_T^i)$ to form the observable gauge boson pairs $(W^i W^i)$.

The leading terms in (19) produce the relation (11) between VEVs. The next-to-leading terms are products between a VEV (EV) and an observable Higgs (gauge) boson. This leads to a linear relation between observable Higgs and gauge bosons:

(20) $\boxed{vH \approx -g^2 \cdot \Sigma_i (\langle W_0^i \rangle W^i)}$      General gauge

     $\sqrt{3}H \approx g \cdot \Sigma_i w^i$      assuming $w^i = w$ for i=1,2,3 and using (11)

The smallest terms in (19) contain products of two observable bosons. Those lead to the relation (14) between observable Higgs and gauge bosons, together with its gauge-invariant generalization (12).

## 4. Symmetry-Breaking Gauge Boson Potential

In order to develop a finite VEV one needs a gauge-invariant potential that has a symmetry-breaking ground state. In the standard model, this is accomplished by the *ad-hoc* potential for the scalar Higgs boson which combines an attractive quadratic term with a repulsive quartic term. These are associated with two adjustable parameters $-\mu^2$ and $\lambda$. The SU(2) gauge bosons, on the other hand, exhibit non-abelian self-interactions which generate a suitable potential dynamically. These do not involve adjustable parameters (apart from the gauge coupling *g* which is also adjustable in the standard model). The gauge boson potential corresponds to the one-loop self-interactions shown in Figure 1. They contain quadratic and quartic terms analogous to the Higgs boson potential. For the diagrams in Fig. 1 these are of O($g^2$) of O($g^4$), respectively. They come with the effective coupling constants $\alpha_0$ and $\alpha_5$. Those can in principle be obtained by evaluating the diagrams in Figure 1, but such a calculation would go beyond the scope of this work. Instead, we will use compatibility with the standard Higgs potential to constrain them.

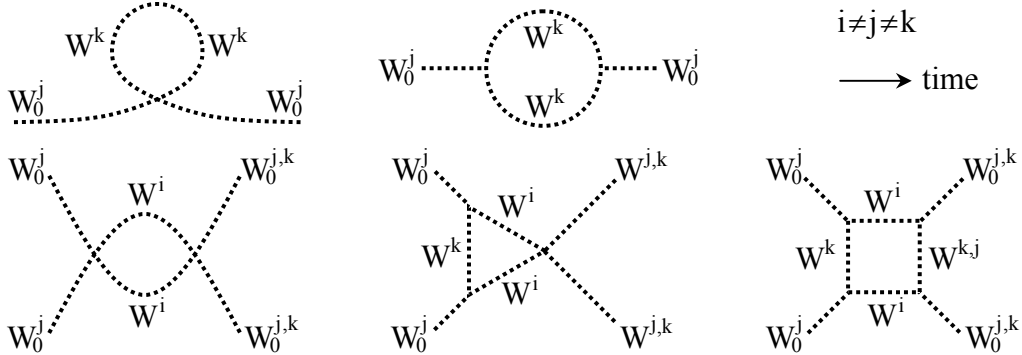

**Figure 1** One-loop self-interactions of the SU(2) gauge bosons. These determine the symmetry-breaking potential. Top row: The quadratic self-energies of O($g^2$) which determine $L_0, \alpha_0$. Bottom row: The quartic + biquadratic self-interactions of O($g^4$) which determine $L_5, \alpha_5$. External lines correspond to Lagrangian gauge bosons $W_0^i$ while internal lines $W^i$ lack a EV. To preserve the equivalence of the three gauge bosons $W_0^i$ they have not been rearranged into $W_0^\pm$ and $W_0^3$.

One can make a generic ansatz for the gauge boson potential (and the corresponding Lagrangian) which satisfies gauge symmetry together with a custodial symmetry [5],[9]. The four-vector $\mathbf{V}_{0,\mu}$ defined in (4)-(6) generates three independent Lagrangians:

(21) $\boxed{L_0 = -\alpha_0 \cdot M_H^2 \cdot \text{tr}[\mathbf{V}_{0,\mu} \mathbf{V}_0^\mu]} \quad \rightarrow \quad \alpha_0 \cdot M_W^2 \cdot \tfrac{1}{2} \Sigma_i (W_0^i W_0^i) \quad M_H^2 = \tfrac{1}{4} v^2 \quad M_W^2 = \tfrac{1}{4} g^2 v^2$

(22) $\boxed{\begin{array}{l} L_4 = \alpha_4 \cdot \text{tr}[\mathbf{V}_{0,\mu} \mathbf{V}_{0,\nu}] \cdot \text{tr}[\mathbf{V}_0^\mu \mathbf{V}_0^\nu] \quad \rightarrow \quad \alpha_4 \cdot \tfrac{1}{4} g^4 \cdot \Sigma_{jk} (W_0^j W_0^k) \cdot (W_0^j W_0^k) \\ L_5 = \alpha_5 \cdot \text{tr}[\mathbf{V}_{0,\mu} \mathbf{V}_0^\mu] \cdot \text{tr}[\mathbf{V}_{0,\nu} \mathbf{V}_0^\nu] \quad \rightarrow \quad \alpha_5 \cdot \tfrac{1}{4} g^4 \cdot \Sigma_{jk} (W_0^j W_0^j) \cdot (W_0^k W_0^k) \end{array}}$

On the right side the Lagrangians have been reduced to the unitary gauge. To keep $\alpha_0$ dimensionless, $L_0$ includes the scale factors $M_H^2$ ($M_W^2$) representing the squared tree-level masses of the Higgs (gauge) bosons [9]. The factors $g^2$ and $g^4$ of the diagrams in Fig. 1 are extracted from the coupling constants $\alpha_0$ and $\alpha_5$. While $L_4$ and $L_5$ consist of quartic + biquadratic terms, their difference $L_{45}$ is purely biquadratic:

(23) $\boxed{L_{45} = \alpha_{45} \cdot (L_5/\alpha_5 - L_4/\alpha_4) \quad \rightarrow \quad \alpha_{45} \cdot \tfrac{1}{4} g^4 \cdot \Sigma_{j \neq k} [(W_0^j W_0^j) \cdot (W_0^k W_0^k) - (W_0^j W_0^k)^2]}$

$L_{45}$ is proportional to the non-abelian gauge Lagrangian $L_{bq}$ which describes the non-abelian vertex between four SU(2) gauge bosons. It does not contribute to the potential of the composite Higgs boson, as explained in Appendix A. That leaves two independent Lagrangians for the dynamical gauge boson potential. These are chosen to be $L_0, L_5$:

(24) $\boxed{V^{dyn} = -(L_0 + L_5) = \alpha_0 \cdot M_H^2 \cdot tr[(V_0 V_0)] - \alpha_5 \cdot \{tr[(V_0 V_0)]\}^2}$ $\qquad M_H^2 = \tfrac{1}{4} v^2$

(25) $\boxed{V^{dyn} = -\alpha_0 \cdot \tfrac{1}{8} g^2 v^2 \cdot \Sigma_i (W_0^i W_0^i) - \alpha_5 \cdot \tfrac{1}{4} g^4 \cdot \{\Sigma_i (W_0^i W_0^i)\}^2}$ $\qquad$ Unitary gauge

Choosing $L_5$ avoids mixed scalar products of the form $(W_0^j W_0^k)$ which would complicate the minimization of the potential. $-L_0$ is an attractive potential arising from the gauge boson self-energies (see [2]). It drives the potential minimum toward a finite EV. The quartic potential $-L_5$ must be repulsive to prevent a runaway of the potential minimum to $-\infty$ at large field amplitudes. To find the appropriate signs for the coupling constants one needs to take into account two minus signs, one from $V = -L$, the other from the negative scalar product of the space-like gauge bosons. Thus, a symmetry-breaking potential with an attractive quadratic term and a repulsive quartic term requires $\alpha_0 < 0$ and $\alpha_5 < 0$.

The gauge boson potential (24) mimics the standard Higgs potential (2). This becomes obvious after replacing the square of the standard Higgs field $\Phi_0^\dagger \Phi_0$ by the square of the composite Higgs field $tr[(V_0 V_0)]$ via the definition (8):

(26) $V_\Phi = -\mu^2 \cdot tr[(V_0 V_0)] + \lambda \cdot \{tr[(V_0 V_0)]\}^2$

The comparison with (24) establishes a simple connection between the Higgs potential parameters $-\mu^2, \lambda$ and the gauge boson couplings $\alpha_0, \alpha_5$:

(27) $\boxed{-\mu^2 = \alpha_0 \cdot M_H^2}$ $\qquad M_H^2 = \tfrac{1}{4} v^2 \qquad \boxed{\lambda = -\alpha_5} \qquad \alpha_0 < 0 , \alpha_5 < 0$

As a consequence of this similarity the minimization of the gauge boson potential becomes similar to that of the scalar Higgs potential. It requires only the solution of a quadratic equation in the variable $\Sigma_i (W_0^i W_0^i)$. The potential takes its minimum over a plane in the three-dimensional space spanned by the coordinates $(W_0^i W_0^i)$, $i = 1, 2, 3$:

(28) $V^{dyn} = 2^{-6} v^4 \cdot \alpha_0^2 / \alpha_5$ $\qquad \boxed{\Sigma_i (\langle W_0^i \rangle \langle W_0^i \rangle) = -\tfrac{1}{4} v^2 / g^2 \cdot \alpha_0 / \alpha_5} \qquad$ at the minimum

This relation connects the EVs of the gauge bosons with the VEV $v$ of the Higgs boson. Assuming equal EVs simplifies $\Sigma_i (\langle W_0^i \rangle \langle W_0^i \rangle)$ to $-3w^2$. This assumption will not be made here to allow for trade-offs between the three EVs $\langle W_0^i \rangle$ allowed by (28).

## 5. Compatibility Criteria and their Consequences

The EVs $\langle W_0^i \rangle$ obtained from the gauge boson potential are connected to the Higgs VEV $v$ by the relations (11),(20). Those originate from the definition (10) of the composite Higgs boson:

(29) $\quad \Sigma_i (\langle W_0^i \rangle \langle W_0^i \rangle) = -v^2/g^2 \qquad\qquad$ General gauge

(30) $\quad \Sigma_i (\langle W_0^i \rangle W^i) \approx -v/g^2 \cdot H \qquad\qquad$ General gauge

Compatibility between the two VEVs obtained in (28) and (29) from the two potentials fixes the ratio of the coupling constants in the gauge boson potential:

(31) $\quad \boxed{\alpha_5/\alpha_0 \approx \tfrac{1}{4}}$

This line of reasoning can be applied to other quantities as well. To obtain a second constraint for $\alpha_0, \alpha_5$ we apply this criterion to the gauge boson mass. In the standard model one obtains $M_W^2 = \tfrac{1}{4} g^2 v^2$ via the gauge-invariant derivatives $D_\mu$ in the kinetic Lagrangian $(D_\mu \Phi_0)^\dagger \cdot (D^\mu \Phi_0)$ of the Higgs boson. This term combines a pair of gauge bosons from $D_\mu$ with a pair of Higgs VEVs $v$ from $\Phi_0$. Attempting a similar scheme with the kinetic Lagrangian of the gauge bosons in (A1),(A2) would not work, since this term vanishes for the gauge bosons that form the composite Higgs boson (see Appendix A). Instead one can use the scheme that determines the Higgs mass in the standard model. After converting the Higgs potential from the Lagrangian field $H_0$ to the observable field $H$, its mass is extracted from the $H^2$ term. Here we convert the gauge boson potential (25) from $W_0^i$ to $W^i$ via (19) and collect the mass terms $(W^i W^i)$:

(32) $\quad (W_0^i W_0^i) \to (\langle W_0^i \rangle \langle W_0^i \rangle) + 2(\langle W_0^i \rangle W^i) + (W^i W^i) \qquad$ Unitary gauge

$\qquad V_M^{dyn} \to -\alpha_0 \cdot \tfrac{1}{8} g^2 v^2 \cdot \Sigma_i (W^i W^i) - \alpha_5 \cdot \tfrac{1}{2} g^4 \cdot \{\Sigma_i (\langle W_0^i \rangle \langle W_0^i \rangle) \cdot \Sigma_i (W^i W^i) + 2 [\Sigma_i (\langle W_0^i \rangle W^i)]^2\}$

$\qquad \approx \{-\alpha_0 \cdot \tfrac{1}{8} g^2 v^2 \cdot \Sigma_i (W^i W^i) - \alpha_5 \cdot \tfrac{1}{2} g^4 \cdot [-v^2/g^2 \cdot \Sigma_i (W^i W^i) + 2 v^2/g^4 \cdot H^2]\}$

$\qquad \approx \underbrace{(\tfrac{1}{4}\alpha_0 - 3\alpha_5)}\, g^2 v^2 \cdot \tfrac{1}{2} \Sigma_i (W^i W^i)$

(33) $\qquad = -M_W^2 \cdot \tfrac{1}{2} \Sigma_i (W^i W^i) \qquad M_W^2 = \tfrac{1}{4} g^2 v^2 \qquad \boxed{(\tfrac{1}{4}\alpha_0 - 3\alpha_5) \approx \tfrac{1}{4}}$

The two sums containing $\langle W_0^i \rangle$ are converted to $v^2, H^2$ via (29),(30). $H^2$ is then converted to $-g^2 \Sigma_i (W^i W^i)$ via (14). The results for $M_W^2$ from the gauge and Higgs boson potentials are compared in (33). The resulting constraint for $\alpha_0, \alpha_5$ is combined with the constraint (31) from the VEVs to obtain the coupling constants of the gauge boson potential:

(34) $\boxed{\alpha_0 \approx -\tfrac{1}{2} \qquad \alpha_5 \approx -\tfrac{1}{8}}$

An evaluation of the diagrams in Fig. 1 can then be used to check the potential (25). The two Lagrangians $L_0, L_5$ are expected to provide the dominant contribution due to their close relation with the standard Higgs potential. But there are additional gauge-invariant Lagrangians available, such as analogs of $L_0, L_5$ which violate custodial symmetry [5].

The ultimate test for this model will require its extension to the full SU(2)×U(1)$_Y$ symmetry of the electroweak interaction, as outlined in Appendix B. It requires evaluations of the complete set of one-loop Feynman diagrams including fermions, Goldstones, gauge-fixing terms, ghosts, and counterterms for renormalization. Related calculations were performed in Refs. [5],[7]-[13]. Most of them considered the high energy limit in order to explore unitarity constraints at the TeV scale.

## 6. Summary and Outlook

A symmetry-breaking mechanism is explored for a SU(2) gauge theory where the gauge bosons themselves break the symmetry. Responsible for that are their quadratic and quartic one-loop self-interactions. The respective coupling constants $\alpha_0, \alpha_5$ can be mapped onto the two parameters $\mu, \lambda$ of the standard Higgs potential via the composite Higgs model proposed in [2]. An estimate of $\alpha_0$ and $\alpha_5$ is obtained by requiring that the observables should not depend on whether they are derived from the gauge bosons or the Higgs boson. That leads to an estimate of the gauge boson self-energies from $\alpha_0$.

A promising result arises when normalizing the relation between the Higgs and gauge fields, including the 4 components of the Higgs field and all 3 polarization states $\alpha$ of the 3 observable gauge fields $W^i$: $[H^2 + \Sigma_i w_i^2] \approx -g^2 \cdot \Sigma_i \Sigma_\alpha (W_\alpha^i W_\alpha^i)$. That determines the value of the weak SU(2) coupling $g$ with tree-level accuracy [3]:

(35) $\quad 4 = g^2 \cdot 3 \cdot 3 \qquad g^2 = 4/9 \qquad \boxed{g = 2/3}$

This minimal SU(2) gauge model may serve as prototype for solving the mass gap problem for Yang-Mills gauge theories in general [4]. It suggests that gauge bosons are able to break their symmetry dynamically and thereby acquire mass. A pair of composite Higgs bosons formed by pairs of gauge bosons represents the simplest possible realization of a glueball in Yang-Mills theories. These objects have been studied extensively for the SU(3) symmetry of the strong interaction [14]. The weaker SU(2) coupling and its smaller group size should make the mass gap problem more tractable. Experimentally, it will be interesting to get access to the threshold $v = 2M_H$ for producing a Higgs pair [15], i.e., the minimal SU(2) glueball in the composite Higgs model.

**Appendix A: Biquadratic Gauge Boson Lagrangians**

While the potential $-(L_0+L_5)$ can be minimized analogous to the standard Higgs potential, the biquadratic Lagrangian $L_{45}$ in (23) introduces scalar products $(W_0^j W_0^k)$ with $j \neq k$ which tend to produce more complex potential surfaces. $L_{45}$ is proportional to the non-abelian part $L_{bq}$ of the kinetic gauge boson Lagrangian:

(A1)  $L_{kin} = -¼ \Sigma_i W_{0\mu\nu}^i W_0^{i\mu\nu}$    $W_{0\mu\nu}^i = [\partial_\mu W_{0\nu}^i - \partial_\nu W_{0\mu}^i] - g \cdot \Sigma_{jk} \varepsilon^{ijk} W_{0\mu}^j W_{0\nu}^k$

(A2)  $L_{bq} = -¼ g^2 \cdot \Sigma_{ijkmn} \varepsilon^{ijk} \varepsilon^{imn} W_{0\mu}^j W_{0\nu}^k \cdot W_0^{m\mu} W_0^{n\nu}$     i,j,k,m,n = 1,2,3

$\phantom{L_{bq}} = -¼ g^2 \cdot \Sigma_{jkmn} [\delta^{jm} \delta^{kn} - \delta^{jn} \delta^{km}] W_{0\mu}^j W_{0\nu}^k \cdot W_0^{m\mu} W_0^{n\nu}$

$\phantom{L_{bq}} = -¼ g^2 \cdot \Sigma_{j \neq k} [(W_0^j W_0^j) \cdot (W_0^k W_0^k) - (W_0^j W_0^k)^2]$

In the last line the diagonal elements j=k vanish, leaving 6 off-diagonal terms with j≠k.

In the following we consider only gauge bosons that contribute to the composite Higgs boson and its potential. These are restricted in (17) to have the same polarization vector $\varepsilon_\alpha$. Gauge bosons can of course exhibit all possible polarization vectors $\varepsilon_\alpha^j, \varepsilon_\alpha^k$ in scattering processes [8].

It is useful to separately consider the two opposing terms in the last line of (A2). Following (17) we decompose Lagrangian gauge bosons into EVs and observable gauge bosons via: $W_0^i = \langle W_0^i \rangle + W_T^i$, $\langle W_0^i \rangle = w \cdot \varepsilon_\alpha$, $W_T^i = w^i \cdot \varepsilon_\alpha$. For mass-like terms of the form $g^2 w^2 \cdot (W^k W^k)$ one finds:

(A3)  $-¼ g^2 \cdot \Sigma_{j \neq k} [(\langle W_0^j \rangle \langle W_0^j \rangle) \cdot (W^k W^k) + (\langle W_0^k \rangle \langle W_0^k \rangle) \cdot (W^j W^j)] =$    1st term

$\phantom{aaaa} = +¼ g^2 w^2 \cdot \Sigma_{j \neq k} [(W^k W^k) + (W^j W^j)] = +g^2 w^2 \cdot \Sigma_i (W^i W^i)$

(A4)  $+¼ g^2 \cdot \Sigma_{j \neq k} [(\langle W_0^j \rangle W^k)^2 + (W^j \langle W_0^k \rangle)^2] = +¼ g^2 w^2 \cdot \Sigma_{j \neq k} [(w^k)^2 + (w^j)^2] =$    2nd term

$\phantom{aaaa} = -¼ g^2 w^2 \cdot \Sigma_{j \neq k} [(W^k W^k) + (W^j W^j)] = -g^2 w^2 \cdot \Sigma_i (W^i W^i)$

In the 2nd term the vectors $\langle W_0^{j,k} \rangle$ or $W^{j,k}$ are first decomposed into $w \cdot \varepsilon_\alpha$ or $w^{j,k} \cdot \varepsilon_\alpha$, then sorted into equal pairs $(w^i)^2$, and eventually converted back into scalar products $(W^i W^i)$. In the last conversion of (A3),(A4) the factor ¼ is compensated by reducing the 12 elements in the sum $\Sigma_{j \ne k}$ to 3 in the sum $\Sigma_i$. The final results for (A3) and (A4) cancel each other. Such cancellations also occur but also for the other terms generated by (A2). This is due to the requirement of equal polarization vectors for the gauge bosons forming the composite Higgs boson which makes it possible to convert the 2$^{nd}$ term to the same form as the 1$^{st}$ term, but with opposite sign. Without this constraint one obtains non-zero mass-like terms with the polarizations $\alpha=\beta$, $\gamma=\delta$ for the 1$^{st}$ term and $\alpha=\gamma$, $\beta=\delta$ for the 2$^{nd}$ term (with $\alpha,\beta,\gamma,\delta$ referring to the gauge bosons j,j,k,k). The Lagrangian $L_{45}$ defined in (25) is proportional to $L_{bq}$ and thus exhibits the same properties.

In addition to the biquadratic vertex (A2) there are other diagrams of O($g^2$) to be considered, as shown in Figure 2. They consist of two trilinear vertices connected by a propagator (compare gauge boson scattering [7],[8]). These vertices originate from mixed products between the derivatives and the non-abelian term in the Lagrangian (A1):

(A5) $\boxed{L_{tri} = \tfrac{1}{2} g \cdot \Sigma_{ijk}\, \varepsilon^{ijk}\, [\partial_\mu W^i_{0\nu} - \partial_\nu W^i_{0\mu}] \cdot W^j_{0\mu} W^k_{0\nu}}$

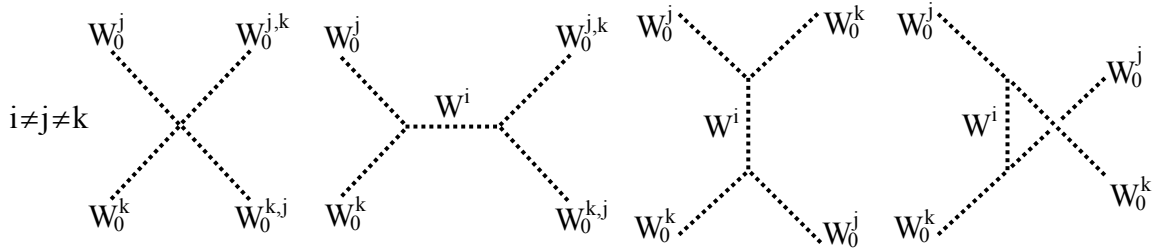

**Figure 2** Biquadratic tree-level interactions of O($g^2$) for SU(2) gauge bosons $W^j_0, W^k_0$ (j≠k). In addition to the quadruple vertex there are three diagrams consisting of trilinear vertices connected by a propagator $W^i$. The EV is eliminated in the propagator by the derivatives in the triple vertices (A5). The equivalence of the three gauge bosons $W^i_0$ has been preserved by avoiding the combination of $W^1_0, W^2_0$ into $W^\pm_0$.

The derivatives associated with the triple vertices eliminate the EV from the internal bosons $W^i$, while the external bosons $W^j_0, W^k_0$ keep their EVs. At the same time, the derivatives introduce momentum-dependent coefficients which can be characterized by the Mandelstam variables s,t,u [7],[8].

**Appendix B:  Extension to the SU(2)×U(1)$_Y$ Electroweak Symmetry**

The electroweak interaction mixes the SU(2) gauge boson $W^3_0$ with the U(1)$_Y$ gauge boson $B_0$ to form the new mass eigenstates $Z_0$ and $A_0$ (the photon). The remaining SU(2) gauge bosons $W^1_0, W^2_0$ form the charge eigenstates $W^\pm_0$:

(B1) $\quad Z_0 = (gW_0^3 - g'B_0)/(g^2+g'^2)^{1/2}$ $\qquad A_0 = (gB_0 + g'W_0^3)/(g^2+g'^2)^{1/2}$

$\quad\quad W_0^+ = (W_0^1 - iW_0^2)/\sqrt{2}$ $\qquad\qquad\qquad W_0^- = (W_0^1 + iW_0^2)/\sqrt{2}$

The ratio $g'/g = \tan\theta_w$ determines the weak mixing angle $\theta_w$. The couplings $g, g'$, and $e$ of the symmetry groups $SU(2), U(1)_Y$, and $U(1)_{EM}$ are given by:

$g/(g^2+g'^2)^{1/2} = \cos\theta_w = c_w \quad g'/(g^2+g'^2)^{1/2} = \sin\theta_w = s_w \quad e = gg'/(g^2+g'^2)^{1/2} = g\cdot\sin\theta_w = g'\cdot\cos\theta_w$

Analogous to (17) one can extract the EVs from the gauge bosons $Z_0$ and $B_0$:

(B2) $\quad Z_0 = \langle Z_0\rangle + Z_T \quad \langle Z_0\rangle = z\cdot\varepsilon_\alpha \quad \langle Z_T\rangle = 0 \quad Z_T = z^i\cdot\varepsilon_\alpha \qquad$ Landau gauge

$\quad\quad B_0 = \langle B_0\rangle + B_T \quad \langle B_0\rangle = b\cdot\varepsilon_\alpha \quad \langle B_T\rangle = 0 \quad B_T = b^i\cdot\varepsilon_\alpha$

The EV of the photon $A_0$ vanishes, since it represents the remaining electromagnetic $U(1)_{EM}$ symmetry. The VEVs $z, b$ are obtained by inserting $\langle W_0^3\rangle = w^3\cdot\varepsilon_\alpha$ from (17) into the EV of (B1), taking into account $\langle A_0\rangle = 0$:

(B3) $\quad \langle B_0\rangle = -g'/g\cdot\langle W_0^3\rangle = -g'/g\cdot w^3\cdot\varepsilon_\alpha \qquad b = -w^3\cdot g'/g = -w^3\cdot s_w/c_w$

$\quad\quad \langle Z_0\rangle = (g^2+g'^2)^{1/2}/g\cdot\langle W_0^3\rangle = z\cdot\varepsilon_\alpha \qquad \boxed{z = w^3\cdot(g^2+g'^2)^{1/2}/g = w^3/c_w}$

The ratio of the VEVs $w^3/z = \cos\theta_w$ is identical to the tree-level mass ratio $M_W/M_Z$.

In order to generalize the definitions of the composite Higgs boson and the symmetry-breaking gauge boson potential to $SU(2)\times U(1)_Y$ one has to include the $U(1)_Y$ gauge boson $B_0$ in the gauge-invariant derivatives (5),(A1):

(B4) $\quad D_\mu U = \partial_\mu U - ig\cdot\mathbf{W}_{0,\mu}U + ig'U\mathbf{B}_{0,\mu} \qquad \mathbf{W}_{0,\mu} = \Sigma_j W_{0,\mu}^j\cdot\tfrac{1}{2}\tau^j \qquad \mathbf{B}_{0,\mu} = B_{0,\mu}\cdot\tfrac{1}{2}\tau^3$

(B5) $\quad W_{0\mu\nu}^i = [\partial_\mu W_{0\nu}^i - \partial_\nu W_{0\mu}^i] - g\cdot\Sigma_{jk}\varepsilon^{ijk}W_{0\mu}^j W_{0\nu}^k \qquad B_{0\mu\nu} = [\partial_\mu B_{0\nu} - \partial_\nu B_{0\mu}]$

From there one can proceed as in Sections 2,3 after the following conversions:

(B6) $\quad W_0^1 \to (W_0^+ + W_0^-)/\sqrt{2} \qquad\qquad\qquad W_0^2 \to i(W_0^+ - W_0^-)/\sqrt{2}$

$\quad\quad W_0^3 \to (gZ_0 + g'A_0)/(g^2+g'^2)^{1/2} = c_w Z_0 + s_w A_0$

The definition (10) of the composite Higgs boson and the relations (11),(20),(14) become:

(B7) $\quad \boxed{H_0^2 = -g^2\cdot[2(W_0^+ W_0^-) + (Z_0 Z_0)/c_w^2]}$ $\qquad\qquad$ Unitary gauge

(B8) $\quad \boxed{v^2 = -g^2\cdot[2(\langle W_0^+\rangle\langle W_0^-\rangle) + (\langle Z_0\rangle\langle Z_0\rangle)/c_w^2]}$ $\qquad$ General gauge

(B9) $\quad \boxed{vH \approx -g^2\cdot[(\langle W_0^+\rangle W^-) + (W^+\langle W_0^-\rangle) + (\langle Z_0\rangle Z_0)/c_w^2]}$ $\quad$ General gauge

(B10) $\quad \boxed{H^2 \approx -g^2\cdot[2(W^+ W^-) + (ZZ)/c_w^2]}$ $\qquad\qquad$ Unitary gauge

The gauge boson potential takes again the form $V^{dyn} = -L^{dyn} = -L_0 - L_5$:

(B11) $\boxed{L_0 = \alpha_0 \cdot M_W^2 \cdot [(W_0^+ W_0^-) + \tfrac{1}{2}(Z_0 Z_0)/c_w^2]}$ $\qquad M_W^2 = \tfrac{1}{4} g^2 v^2$

$\boxed{L_5 = \alpha_5 \cdot \tfrac{1}{4} g^4 \cdot [(W_0^+ W_0^-)^2 + (W_0^+ W_0^-)(Z_0 Z_0)/c_w^2 + \tfrac{1}{4}(Z_0 Z_0)^2/c_w^4]}$

The photon is massless and therefore does not contribute to $L_0, L_5$ which are built from mass Lagrangians. The potential minimum has the same depth as in (28) for pure SU(2):

(B12) $V^{dyn} = 2^{-6} v^4 \cdot \alpha_0^2/\alpha_5 \quad$ for $\quad [2(\langle W_0^+\rangle\langle W_0^-\rangle) + (\langle Z_0\rangle\langle Z_0\rangle)/c_w^2] = -\tfrac{1}{4} v^2/g^2 \cdot \alpha_0/\alpha_5$

The minimum extends now along a line in the two-dimensional space spanned by the coordinates $(W_0^+ W_0^-)$ and $(Z_0 Z_0)$, as shown in Fig. 7a of [2]. A well-defined point on this line can be selected by requiring identical VEVs $w^i$ for the three gauge bosons $W_0^j$, even after mixing. This converts (B8) into $v^2 = g^2 [2w^2 + z^2/c_w^2]$ and (B3) into $z^2 = w^2/c_w^2$, with the common VEV $w^i = w = v/[g(2+c_w^{-4})^{1/2}] = 194$ GeV and $z = 220$ GeV. The ratio of the VEVs $w/z = \cos\theta_w$ then becomes equal to the tree-level mass ratio $M_W/M_Z$.

The compatibility criteria (31),(33) for the coupling constants $\alpha_0, \alpha_5$ require identical results for gauge boson EVs and masses from either the Higgs potential (2) or the gauge boson potential (B11). The comparison of the EVs in (B8) and (B12) yields:

(B13) $\boxed{\alpha_5/\alpha_0 = \tfrac{1}{4}}$

This holds for all values of $g'$. Consistency of the gauge boson masses involves the quadratic part of the potential for observable gauge bosons. As in (32),(33) one obtains:

(B14) $(W_0^+ W_0^-) \rightarrow (\langle W_0^+\rangle\langle W_0^-\rangle) + [(\langle W_0^+\rangle W^-) + (W^+\langle W_0^-\rangle)] + (W^+ W^-)$

$(Z_0 Z_0) \rightarrow (\langle Z_0\rangle\langle Z_0\rangle) + 2(\langle Z_0\rangle Z_0) + (ZZ)$

$V_M^{dyn} \rightarrow -\alpha_0 \cdot \tfrac{1}{8} g^2 v^2 \cdot [2(W^+ W^-) + (ZZ)/c_w^2] \qquad\qquad$ Unitary gauge

$\qquad -\alpha_5 \cdot \tfrac{1}{2} g^4 \cdot \{[2(\langle W_0^+\rangle\langle W_0^-\rangle) + (\langle Z_0\rangle\langle Z_0\rangle)/c_w^2] \cdot [2(W^+ W^-) + (ZZ)/c_w^2]$

$\qquad\qquad + 2[(\langle W_0^+\rangle W^-) + (W^+\langle W_0^-\rangle) + (\langle Z_0\rangle Z_0)/c_w^2]^2\}$

$\qquad \approx (\tfrac{1}{4}\alpha_0 - 3\alpha_5) \cdot g^2 v^2 \cdot [(W^+ W^-) + \tfrac{1}{2}(ZZ)/c_w^2] = -[M_W^2 \cdot (W^+ W^-) + \tfrac{1}{2} M_Z^2 \cdot (ZZ)]$

Use of (B8)-(B10) leads to the last line. The compatibility condition becomes:

(B15) $M_W^2 = \tfrac{1}{4} g^2 v^2 \qquad M_Z^2 = M_W^2/c_w^2 \qquad \boxed{(\tfrac{1}{4}\alpha_0 - 3\alpha_5) \approx \tfrac{1}{4}}$

This is again independent of $g'$, leaving the combined result (34) for $\alpha_0, \alpha_5$ unchanged:

(B16) $\boxed{\alpha_0 \approx -\tfrac{1}{2} \qquad \alpha_5 \approx -\tfrac{1}{8}}$

The full set of diagrams for the Lagrangians $L_0, L_5$ (including fermions) is given in Fig. 4 of [2]. Calculations related to the coupling constants $\alpha_0, \alpha_5$ have been published in [5],[7]-[13]. They still need to be performed for an energy scale comparable to the EVs $\{w,z\}\cdot\varepsilon_\alpha$ around which the fields W,Z oscillate. One can estimate the (transverse) self-energy $\Sigma_T$ of the gauge bosons using the identification $L_0 \approx [\Sigma_T^W \cdot (W_0^+ W_0^-) + \Sigma_T^Z \cdot \tfrac{1}{2}(Z_0 Z_0)]$, i.e., replacing $\{M_W^2, M_W^2/c_w^2\}$ by $\{\Sigma_T^W, \Sigma_T^Z\}$ in (B11). Assuming a common VEV $w^i = w$ we can use the values of $\{w,z\}$ from above to obtain:

(B17) $\quad \Sigma_T^W(w^2) \approx \Sigma_T^Z(z^2) \approx -\alpha_0 \cdot \tfrac{1}{4} g^2 v^2 \approx -(57.7\,\text{GeV})^2 \quad$ at $\quad w=194\,\text{GeV},\ z=220\,\text{GeV}$